\documentstyle[epsfig,cite]{mn2e}

\begin{document}

\title[{\sl HST}/WFPC2 snapshot imaging of symbiotic stars]
    {{\sl HST}/WFPC2 snapshot imaging of symbiotic stars}
\author[Brocksopp et al.]
    {C.~Brocksopp$^{1,2}$\thanks{email: cb4@mssl.ucl.ac.uk}, M.F.~Bode$^1$, S.P.S.~Eyres$^3$\\
$^1$Astrophysics Research Institute, Liverpool John Moores University, Twelve Quays House, Egerton Wharf, Birkenhead CH41 1LD\\
$^2$Mullard Space Science Laboratory, University College London, Holmbury St. Mary, Dorking, Surrey RH5 6NT\\
$^3$Centre for Astrophysics, University of Central Lancashire, Preston PR1 2HE\\}

\date{Accepted ??. Received ??}
\pagerange{\pageref{firstpage}--\pageref{lastpage}}
\pubyear{??}
\maketitle

\begin{abstract}
The results of a {\sl HST}/WFPC2 snapshot imaging survey of selected symbiotic stars in 1999/2000 are presented. Seven sources -- HD~149427 (PC 11), PU Vul, RT Ser, He2$-$104 (Southern Crab), V1329 Cyg (HBV 475), V417 Cen, AS 201 -- were observed in filters F218W (ultraviolet continuum), F502N ([{\sc Oiii}]$\lambda\lambda\,\, 4959,5007$) and F656N ({\sc H}$\alpha\lambda\,\, 6563$); an eighth source, RS Oph, was observed in F437N ([{\sc Oiii}]$\lambda\,\, 4363$), F502N and F656N. The presence of extended emission was detected in He2$-$104, V1329 Cyg and possibly HD~149427. In He2$-$104 we detected the [{\sc Oiii}] and {\sc H}$\alpha$ counterparts to the inner lobes found in [{\sc Nii}] by Corradi et al. For V1329 Cyg, comparison with previously published {\sl HST}/FOC results indicates expanding ejecta which may be associated with an ejection event in 1982 ($\pm2$ years) at a velocity of $260\pm 50$ km~s$^{-1}$ in the plane of the sky and at an assumed distance of 3.4kpc. We also present previously unpublished radio images of HD~149427, which we have obtained from the archives of the Australia Telescope Compact Array and which reveal the presence of extended emission at a similar orientation to that of the possible optical extension. Finally we also include {\sl HST}/WFPC2 GO observations of AG Peg and detect possible extended emission in the F218W filter. 

\end{abstract}


\section{Introduction}

The current availability of high spatial-resolution telescopes at both optical and radio wavelengths has the potential to revolutionise the field of symbiotic star research. In particular observations with the Hubble Space Telescope ({\sl HST}), the Multi-Element Radio Linked Interferometer Network (MERLIN) and the Very Large Array (VLA) have revealed extended structure to a number of symbiotic systems, providing insight to their nature and physical properties which could not have been determined otherwise (Bode 2003).

Our imaging campaign has previously concentrated on {\sl HST}/WFPC2, MERLIN and VLA imaging of V1016 Cyg (Brocksopp et al. 2002, Watson et al. 2000), HM Sge (Eyres et al. 1995, 2001; Richards et al. 1999) and CH Cyg (Crocker et al. 2001, 2002; Eyres et al. 2002); these observations have enabled the structure of the nebula to be probed down to sub-arcsecond scales, as well as the determination of nebular densities and temperatures, the variation of optical extinction across the nebula and binary parameters. Particularly interesting are the apparently precessing radio jet in CH Cyg and the discovery of non-thermal radio emission in CH Cyg and HM Sge. Brocksopp et al. (2002) also used {\sl HST}/STIS spectra to spatially resolve the nebula of V1016 Cyg and found that it extends for 10--15 arcsec in the ultraviolet, thereby confirming the presence of the nebula first found in the optical by Bang et al. (1992).

A number of other imaging projects have taken place, most notably in the optical by Corradi et al. (1999, 2001 and references therein). Through these optical studies fifty one symbiotic stars were observed in 1998, fourteen of which were resolved and some were found to be highly extended (e.g. He2$-$104, BI Cru). Kenny (1995) also observed He2$-$106, BI Cru, RR Tel, HD~149427, Z And, AG Peg, HM Sge and V1016 Cyg in the radio, all of which were resolved at arcsecond spatial scales with the probable exception of BI Cru.

We have now extended our survey to nine additional sources; AG Peg was part of the same {\sl HST} GO programme in which HM Sge, CH Cyg and V1016 Cyg were observed. A further eight sources -- HD~149427 (PC 11), PU Vul, RT Ser, He2$-$104 (Southern Crab), V1329 Cyg (HBV 475), V417 Cen, AS 201 and RS Oph -- were also observed in an {\sl HST}/WFPC2 snapshot programme. The complete target list for this programme comprised symbiotics showing extended radio emission and/or evidence of high velocity outflows. Previous observations of each source at a range of wavelengths are briefly summarised in the remainder of this section; Section 2 describes the observations, Section 3 presents and discusses our results and we then draw our conclusions in Section 4.

\begin{table}
\caption{Properties of the filters used in our {\sl HST}/WFPC2 observations.} 
\begin{tabular}{lcccc}
\hline
\hline
Filter&Peak $\lambda$&$\Delta\lambda$&Scientific Features\\
&(\AA)&(\AA)&(\AA)&\\
\hline
F218W          & 2091&355.9& Hot continuum\\
F437N          & 4368&25.2&   [O {\sc iii}] $\lambda$4363\\
F469N          & 4699&24.9&   He {\sc ii} $\lambda$4686\\
F487N          & 4863&25.8&   H$\beta$ $\lambda$4861\\
F502N          & 5009&26.8&   [O {\sc iii}] $\lambda\lambda$4959,5007\\
F547M          & 5362&486.6& Cool continuum\\
F656N          & 6561&22.0&   H$\alpha$ $\lambda$6563\\
\hline
\end{tabular}
\label{agpeg}
\end{table}

\begin{table*}
\caption{List of the eight snapshot targets and details of the observations, peak and $1\sigma$ background fluxes in each image and properties of the observed central sources. S, D, D\arcmin=S-, D- and D\arcmin-type symbiotic, SN=symbiotic nova, RN= recurrent nova, YS=yellow symbiotic, PN=planetary nebula (see text for references).}
\begin{tabular}{lcllcccc}
\hline
\hline
Source&Symbiotic&Date of & WFPC2&Total Exposure&Peak Flux&$1\sigma$ Background Level&Central\\
&Type&Observations&Filter&Time (s)& \multicolumn{2}{c}{(erg~s$^{-1}$~cm$^{-2}$~\AA$^{-1}$~arcsec$^{-2}$)}&Source\\
\hline
HD~149427&YS,D\arcmin/PN?&1999 Jul 24&   F218W&20  &$1.77\times 10^{-13}$&$2.00\times 10^{-14}$&Point source\\
(PC 11) &&          &   F502N         &280 &$3.84\times 10^{-12}$&$6.82\times 10^{-16}$&Extended?\\
        &&          &   F656N         &200 &$2.66\times 10^{-12}$&$4.47\times 10^{-16}$&Extended?\\
PU Vul&S,SN&        1999 Jul 30&     F218W       &20 &$4.18\times 10^{-12}$&$1.95\times 10^{-14}$&Point source\\
             &&      &  F502N         &280 &$3.78\times 10^{-12}$&$9.30\times 10^{-16}$&Saturated\\
             &&     &   F656N         &200 &$2.70\times 10^{-12}$&$2.19\times 10^{-15}$&Saturated\\
RT Ser &S,SN&        1999 Aug 26&    F218W       &20 &Non-detection        &$1.99\times 10^{-14}$&Non-detection\\
              &&    &   F502N                  &280 &$2.43\times 10^{-13}$&$4.57\times 10^{-16}$&Point source\\
               &&   &   F656N                  &200 &$2.67\times 10^{-12}$&$3.72\times 10^{-16}$&Saturated\\
He2$-$104&D        &1999 Sep 28&   F218W        &20 &$1.73\times 10^{-13}$&$1.95\times 10^{-14}$&Point source\\
(Southern&&         &   F502N         &280 &$3.96\times 10^{-12}$&$5.04\times 10^{-16}$&Extended\\
Crab)&&             &   F656N             &200 &$2.68\times 10^{-12}$&$5.80\times 10^{-16}$&Extended\\
V1329 Cyg&S,SN&      1999 Oct 19&    F218W       &20 &$1.66\times 10^{-13}$&$1.96\times 10^{-14}$&Point source\\
(HBV 475)&&         &   F502N         &280 &$1.14\times 10^{-12}$&$6.13\times 10^{-16}$&Extended\\
&&                  &   F656N                  &200 &$2.69\times 10^{-12}$&$4.12\times 10^{-16}$&Extended\\
V417 Cen&D\arcmin?, YS&  1999 Oct 22&    F218W   &20 &Non-detection       &$2.66\times 10^{-14}$&Non-detection\\
&&                  &   F502N                  &280 &$3.92\times 10^{-12}$&$6.33\times 10^{-16}$&Point source\\
&&                  &   F656N                  &200 &$2.80\times 10^{-12}$&$4.08\times 10^{-16}$&Pointing error\\
AS 201&D\arcmin     &    1999 Oct 28&    F218W          &20 &$5.22\times 10^{-13}$&$1.98\times 10^{-14}$&Point source\\
&&                   &  F502N                  &280 &$3.79\times 10^{-12}$&$5.44\times 10^{-16}$&Point source\\
&&                  &   F656N                  &200 &$2.65\times 10^{-12}$&$7.29\times 10^{-16}$&Saturated\\
RS Oph&S,RN&       2000 Jun 12&      F437N          &200 &$4.23\times 10^{-12}$&$1.41\times 10^{-15}$&Point source\\
&&                 &    F502N                  &200 &$3.79\times 10^{-12}$&$6.10\times 10^{-16}$&Point source\\
&&                 &    F656N                  &100 &$5.26\times 10^{-12}$&$7.98\times 10^{-16}$&Saturated\\
\hline
\end{tabular}
\begin{tabular}{l}
NB. Those images labelled `saturated' have noticeably flat-topped point spread functions and display `bleeding' in the vertical direction\\ along the CCD. However, the similarity between the peak flux values of a number of the point sources would suggest that some of these\\ are also (close to) saturated in the most central regions.\\
\hline
\end{tabular}
\label{values}
\end{table*}

\section{Observations}

\subsection{Optical}

The {\sl HST}/WFPC2 was used to observe each of our eight SNAP targets (programme 8332) in 1999/2000. The F218W, F502N and F656N filters were used in order to explore the UV continuum, [{\sc Oiii}] $\lambda\lambda4959,5007$ and {\sc H}$\alpha\,\,\lambda6563$ regions of the spectrum respectively (see Tables 1 and 2). The only exception to this was RS Oph which was observed in the F437N ([{\sc Oiii}] $\lambda4363$) filter instead of F218W. Each image was obtained using the Planetary Camera (PC) chip, the spatial resolution of which is 0.045\arcsec~pixel$^{-1}$, with the exception of V417 Cen which was detected by the WF4 chip (due to positional error) which has a spatial resolution of 0.1\arcsec~pixel$^{-1}$. 

Two exposures were obtained in each case and total exposure times for each source were 20, 280 and 200 seconds for the F218W, F502N and F656N filters respectively. Total exposure times for RS Oph were 200, 200 and 100 seconds for F437N, F502N and F656N respectively. Each pair of images was combined and cosmic ray rejected using {\sc crreject} in {\sc iraf}. 

\begin{figure*}
\begin{center}
\leavevmode
\psfig{file=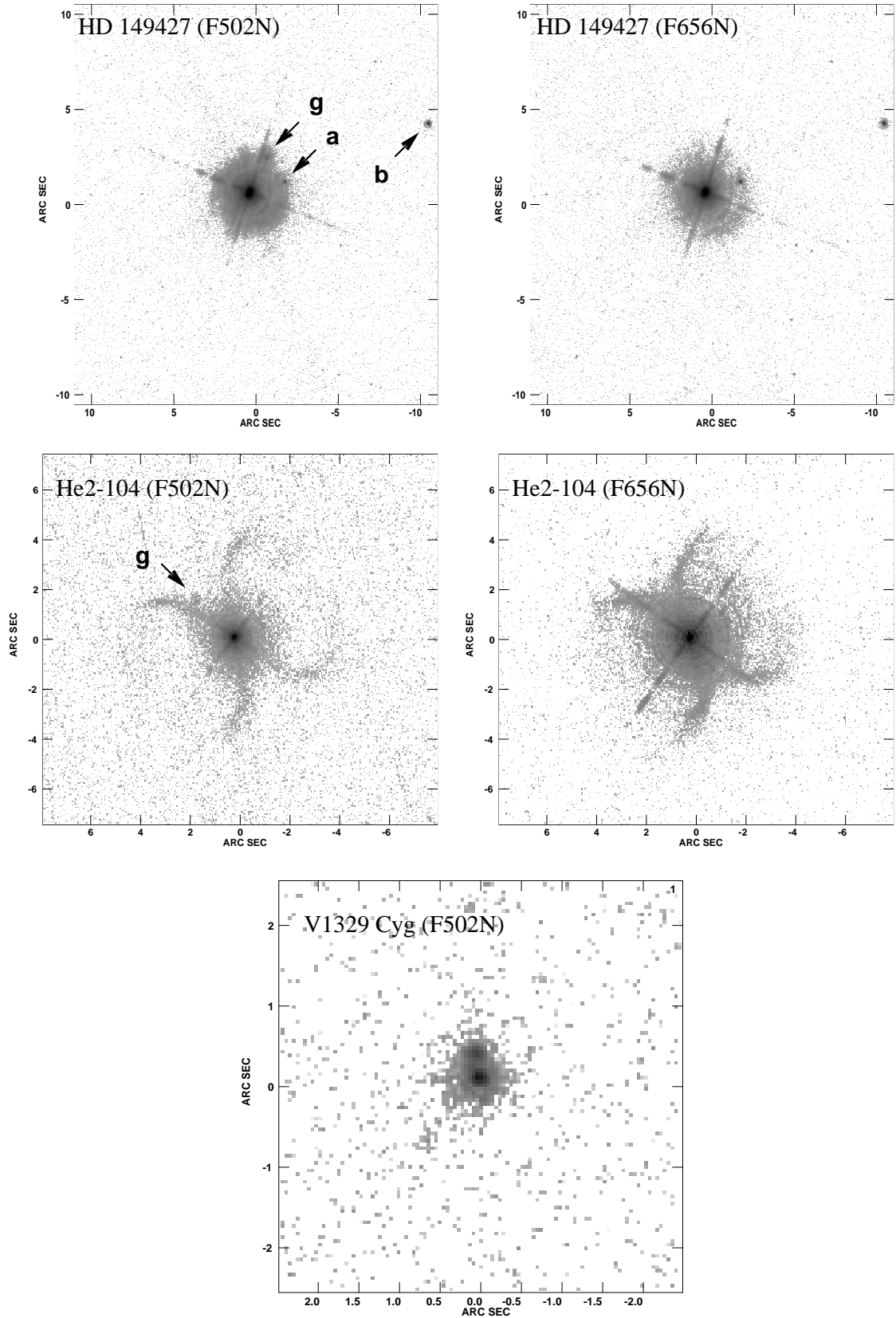,width=15cm}
\caption{{\sl HST}/WFPC2 snapshot images of HD~149427 (top), He2$-$104 (middle) and V1329 Cyg (bottom). Each source appears to show some form of extension. The small arrows and labels refer to the two additional points of emission ({\bf a}, {\bf b}) which may be associated with HD~149427 and the F502N filter ghost ({\bf g}). North is to the top and West to the right in each case.}
\label{snapshots} 
\end{center}
\end{figure*}
 
We also include here images of AG Peg obtained on 2000 April 15 as part of our GO programme 8330; this same programme also observed V1016 Cyg, HM Sge and CH Cyg, the results of which are published elsewhere (Brocksopp et al. 2002; Eyres et al. 2001, 2002; Crocker et al. 2001, 2002; Skopal et al. 2002). Filters F218N, F437N, F469N, F487M, F502N, F547N and F656N were used and further details of exposure time, peak and background flux values and spectral region are listed in Table~\ref{agpeg}. As with the other GO targets, AG Peg was observed during two exposures in each filter, using the PC chip in both normal and dithered modes wth spatial resolutions of 0.045 and 0.023\arcsec~pixel$^{-1}$ respectively. Each pair of normal images was combined and cosmic ray rejected as above. The dithered images were cosmic ray rejected using a mask and then combined using the {\sc drizzle} software within {\sc iraf}.

\subsection{Radio}

Radio observations of HD~149427, which had taken place on 1991 October 31 and 1996 April 3, were obtained from the archive of the Australia Telescope Compact Array (ATCA). These data are previously unpublished, although the first epoch of observations has been analysed by Kenny (1995). All observations made use of the full 6km array which gives an angular resolution of $\sim1\arcsec$ at 8.640 GHz.

The data have been (re-)reduced using standard flagging, calibration and imaging (with natural weighting) techniques within {\sc miriad}; the flux calibrator was PKS 1934$-$638 and the phase reference source was PKS J1723-6500 (NGC 6328) for both epochs. On-source integration time for the 1991 epoch was 8.5 hours at 8.64 GHz; in 1995 data were obtained at 1.344, 2.382, 5.056, 8.512 and 9.024 GHz for 0.85, 0.85, 0.5, 8.53 and 8.03 hours respectively.

\section{Results and discussion}

\subsection{Snapshot sources}

All eight sources were detected through the F502N and F656N filters and all but RT Ser and V417 Cen were detected in the F218W observation (as noted above, RS Oph which was observed and detected through the F437N filter instead of F218W); details of the peak and background fluxes (in units of erg~s$^{-1}$~cm$^{-2}$~\AA$^{-1}$~arcsec$^{-2}$) of each image can be found in Table~\ref{values}.

Images of HD~149427, He2$-$104 and V1329 Cyg are shown in Fig.~\ref{snapshots} and have been chosen by virtue of various noteworthy features. The HD~149427 images are saturated but two small additional peaks of emission are detected in both [{\sc Oiii}] and H$\alpha$ filters (indicated with arrows and labelled {\bf a} and {\bf b} in the F502N image); it is possible that this emission is associated with the source (see below), although additional images and proper motion measurements would be necessary to confirm this. The `hour glass' structures at the centre of the [{\sc Nii}] images of He2$-$104 (Corradi et al. 2001) are also seen here in the [{\sc Oiii}] and H$\alpha$ filters, although our images are insufficiently deep to detect the (presumed) complete ring. V1329 Cyg also shows an additional peak of emission, as observed previously by Schild \& Schmid (1997) and we again discuss this further below.

\begin{figure}
\begin{center}
\leavevmode
\psfig{file=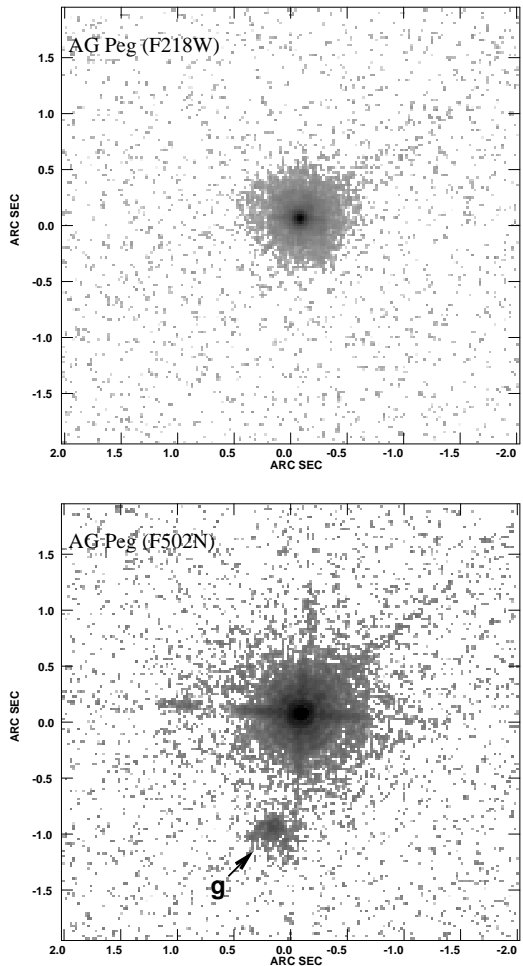,angle=0,width=7cm}
\caption{Dithered images (spatial resolution 0.0223\arcsec) of AG Peg in the F218W (top) and F502N (bottom) filters. The F218W image appears to have some degree of extended emission around the central source, particularly in the horizontal direction of the CCD. The F502N image is saturated but shows clearly the filter ghost ({\bf g}) that is present in the majority of our F502N images. North is up.}
\label{agpegimages} 
\end{center}
\end{figure}

All images not displayed here were either saturated (as in the case of almost all F656N images) or showed no significant deviation from a point source when compared with the output from {\sc tiny tim} (see below). Further details are included in Table~\ref{values}. We note the presence of a filter ghost in each of the more saturated F502N images, just below the bottom right (prior to rotation of the image) diffraction spike; this is shown very clearly in the F502N image of AG Peg (Fig.~\ref{agpegimages} and we suspect that the `jet' in the HD~149427 image discussed by Parthasarathy et al. (2000) was probably related to this ghost -- see feature {\bf g} in Fig.~\ref{snapshots}).

\begin{figure}
\begin{center}
\leavevmode
\psfig{file=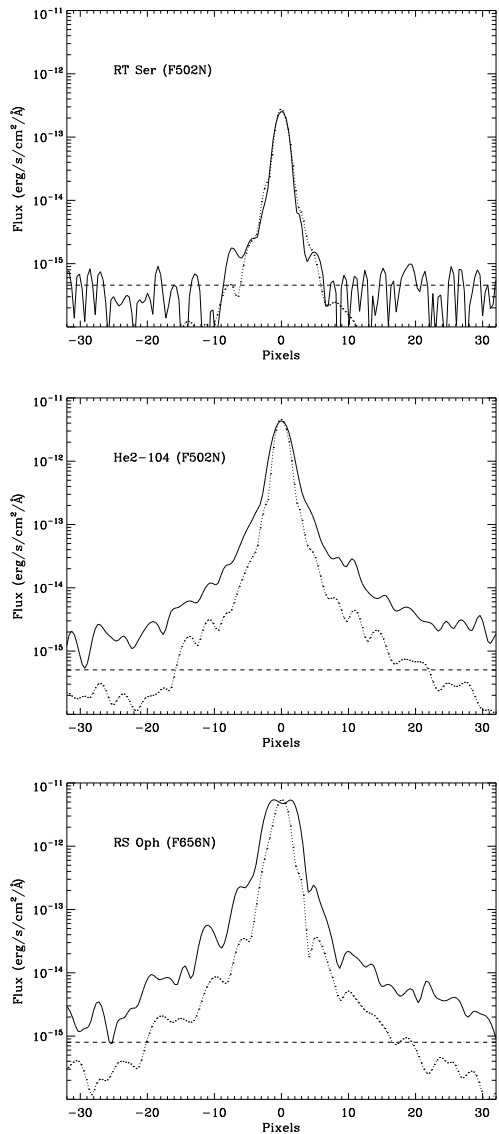,angle=0,width=7cm}
\caption{One-dimensional slices through three illustrative images compared with the equivalent one-dimensional slice through a model image as produced by the {\sc tiny tim} software. Top: F502N image of RT Ser is comparable with a stellar model. Middle: F502N image of He2$-$104 shows excess emission compared with an unresolved stellar model. Bottom: F656N image of RS Oph shows saturation. The horizontal line in each plot indicates the $1\sigma$ background level in the corresponding images.}
\label{slices} 
\end{center}
\end{figure}

The {\sc tiny tim} software was used in order to generate model images of point sources in each filter for comparison with the symbiotic sources. One-dimensional slices of width 1 pixel through each of the WFPC2 images and the model images were obtained at 18 azimuthal angles (chosen so as to exclude those azimuthal angles which included a diffraction spike) using the {\sc slice} routine within {\sc aips}. Plotting each source slice on the same axes as its corresponding model slice indicated clearly which were point sources, which were saturated and which showed structure in terms of excess emission. Selected plots are shown in Fig.~\ref{slices}.

\begin{figure}
\begin{center}
\leavevmode
\psfig{file=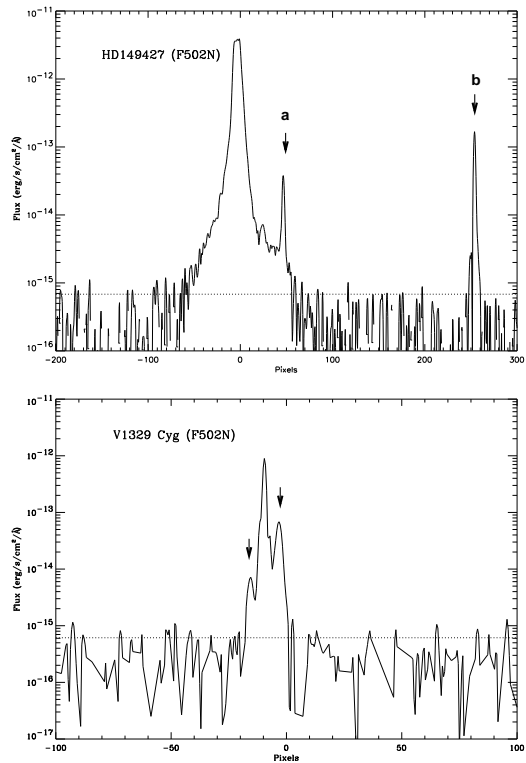,angle=0,width=7cm}
\caption{Top: One-dimensional slice through the F502N image of HD~149427. The peaks related to the two `blobs' of emission are directly in line with the peak of the central object. A slice through the F656N image also intersects the three peaks. Bottom: One-dimensional slice through the F502N image of V1329 Cyg, again passing directly through the peaks of the two components shown in the image (and possibly also through a third and fainter peak as indicated by the arrow). The horizontal line in each plot indicates the $1\sigma$ background level in the corresponding images.}
\label{jet}
\end{center}
\end{figure}

The top plot in Fig.~\ref{slices} indicates that the F502N image of RT Ser shows no significant deviation from the point source model image; this was also the case for all F218W images, the F437N image of RS Oph, the F502N images of AS 201, RS Oph and V417 Cen and the F656N image of V417 Cen. In contrast to this the middle plot shows that there is excess emission, which is attributable to extended nebular material surrounding He2$-$104, in addition to the rings seen in the images -- we note that this excess emission is seen at all azimuthal angles and appears to be symmetrical about the central object, consistent with the model proposed for the inner lobes by Corradi et al. (2001). The bottom plot is included for completeness and shows the F656N profile of RS Oph, one of the saturated images.

\begin{table}
\caption{List of the filters, peak and $1\sigma$ background fluxes and the exposure times used in our {\sl HST}/WFPC2 observations of AG Peg.} 
\begin{tabular}{lccc}
\hline
\hline
Filter&Peak Flux&$1\sigma$ Background Level&Total Exp.\\
& \multicolumn{2}{c}{(erg~s$^{-1}$~cm$^{-2}$~\AA$^{-1}$~arcsec$^{-2}$)}&Time (s)\\
\hline
F218W$^\dagger$&$2.27\times 10^{-10}$&$4.39\times 10^{-14}$&40 \\   
F218W          &$3.62\times 10^{-11}$&$5.86\times 10^{-15}$&100  \\ 
F437N$^\dagger$&$4.93\times 10^{-11}$&$7.02\times 10^{-15}$&200   \\
F437N          &$2.48\times 10^{-12}$&$5.42\times 10^{-16}$&1000  \\
F469N$^\dagger$&$6.79\times 10^{-11}$&$1.26\times 10^{-14}$&80    \\
F469N          &$3.57\times 10^{-12}$&$7.25\times 10^{-16}$&520   \\
F487N$^\dagger$&$2.72\times 10^{-10}$&$3.89\times 10^{-14}$&20    \\
F502N$^\dagger$&$1.04\times 10^{-10}$&$1.31\times 10^{-14}$&40    \\
F502N          &$1.05\times 10^{-11}$&$1.80\times 10^{-15}$&100   \\
F547M$^\dagger$&$1.06\times 10^{-11}$&$1.22\times 10^{-15}$&10    \\
F656N$^\dagger$&$2.07\times 10^{-11}$&$2.57\times 10^{-14}$&10    \\
F656N          &$5.46\times 10^{-13}$&$9.51\times 10^{-16}$&100   \\
\hline
\end{tabular}
\begin{tabular}{l}
\hspace*{0cm}$\dagger$ image observed in the dithered mode in order to improve\\
\hspace*{0.25cm}the spatial resolution\\
\end{tabular}
\label{agpeg}
\end{table}

\subsubsection{HD~149427}

\begin{figure}
\begin{center}
\leavevmode
\psfig{file=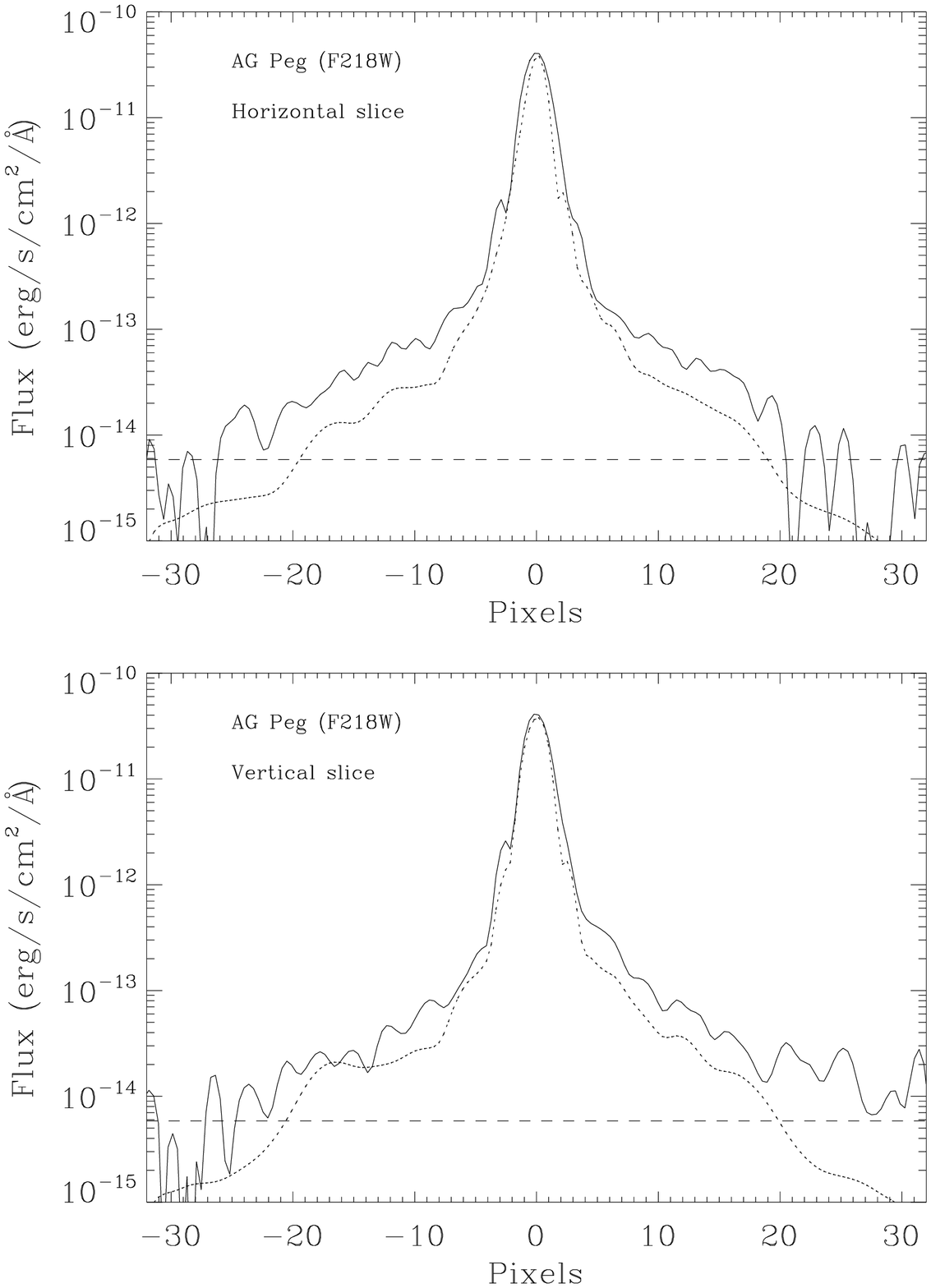,angle=0,width=7cm}
\caption{Profile of AG Peg in the horizontal and vertical directions of the CCD (using the non-dithered image so as to compare with the outout from {\sc tiny tim}). There appears to be a small degree of faint extended emission around the source, particularly in the horizontal direction. The horizontal line in each plot indicates the $1\sigma$ background level in the corresponding images.}
\label{agpegslice} 
\end{center}
\end{figure}

The top plot of Fig.~\ref{jet} shows a similar slice through the F502N image of HD~149427 and illustrates that the three peaks in the image are aligned to within a pixel of each other; this suggests that the emission peaks in HD~149427 may be associated with the central object. Assuming a distance of 5.4 kpc (Kenny 1995) the separation of the two emission peaks are $\sim 12000$ and 64000 AU for the inner ({\bf a}) and outer ({\bf b}) peaks respectively. We acknowledge that this may instead be a chance alignment of stars on the field -- indeed, emission point {\bf b} can be seen in the DSS2-Red catalogue (R.M. Corradi, private communication). However we also note that evidence for jet-like [{\sc Oiii}] emission moving away from the central nebula at $\sim 120$ km~s$^{-1}$ has been discovered via optical spectroscopy (Guti\'errez-Moreno \& Moreno 1998) and so we suggest that these peaks are worthy of further study (see also Section 3.3). While the asymmetry of the possible ejecta also makes an association seem unlikely, such apparently one-sided ejecta have been observed in other sources (e.g. CH Cyg; Crocker et al. 2001). Additional observations would be extremely useful in order to confirm the nature of features {\bf a} and {\bf b} via searches for proper motion and/or spectroscopy. Subsequently, if they do turn out to be ejecta, it would be important to determine whether these peaks are discrete ejecta or bright knots in an underlying faint jet.

\subsubsection{V1329 Cyg}

Likewise the bottom plot of Fig.~\ref{jet} shows a slice through the two peaks in the V1329 Cyg F502N image. We find that they are 280 mas apart, equivalent to $950\pm50$ AU if we assume a distance of 3.4 kpc (M\"urset \& Nussbaumer 1994). 

We compare this with the 1991 {\sl HST}/FOC image of Schild \& Schmid (1997) and find that the lines joining the two respective peaks of each image lie at the same position angle; the separation in the earlier image was 500 AU and consequently we calculate a velocity of $\sim260\pm 50$ km~s$^{-1}$ in the plane of the sky (assuming an error of $\sim10\%$ on the 1991 separation). This in turn would suggest that this mass ejection was not associated with the 1964 outburst as suggested by Schild \& Schmidt (1997), but took place instead in $\sim 1982$ ($\pm 2$ years) and thus was probably associated with the peculiar brightening seen in the optical photometry and the variability of the {\sc Oiii} $\lambda$5007 line during the mid-1980's (Chocol et al. 1999; Wallerstein et al. 1989).

Furthermore Schild \& Schmid (1997) determined an orbital inclination angle of $86\pm 2^{\circ}$, suggesting that the mass ejection occurred {\em along} the orbital plane. Either this source is highly unusual in that it is ejecting material in the plane of the orbit, as opposed to perpendicularly to the orbit as in most other sources, or the spectropolarimtery used to determine the orbit was affected by the presence of the ejected material along this axis.

\subsection{AG Peg}

As with a number of the snapshot sources, all of the AG Peg images were saturated with the exception of those taken through the ultraviolet filter; this was the case for both dithered and non-dithered exposures. The dithered F218W and F502N images are displayed in Fig.~\ref{agpegimages}, the latter included so as to show the filter ghost that is present in most of the F502N images mentioned in this paper.

A slice through the non-dithered ultraviolet image, obtained and compared with the {\sc tiny tim} output as previously, shows that there may be a small degree of faint extended emission around the central source detected to a radius of $\sim20$ pixels ($\sim0.9$ arcsec).

\begin{table}
\caption{Peak and integrated flux densities of each of the radio observations of HD~149427.}
\begin{tabular}{lcccc}
\hline
Date& Frequency& Peak Flux& Int. Flux&$\sigma$\\
&&Density&Density&\\
&(GHz)& (mJy~beam$^{-1}$)&(mJy)& (mJy)\\
\hline
1991 Oct 31&8.640&10.60&16.02&0.43\\
1996 April 3&1.344&--&--&0.36\\
1996 April 3&2.382&2.7$^*$&--&0.28\\
1996 April 3&5.056&13.36&15.27&0.27\\
1996 April 3&8.512&23.48&25.73&0.10\\
1996 April 3&9.024&24.53&26.30&0.10\\
\hline
\end{tabular}
\begin{tabular}{l}
*This value was obtained via measurement of the peak pixel value\\ since there was insufficient data for a Gaussian fit.
\end{tabular}
\label{radio}
\end{table}

\subsection{Radio images of HD~149427}
\begin{figure*}
\begin{center}
\leavevmode
\psfig{file=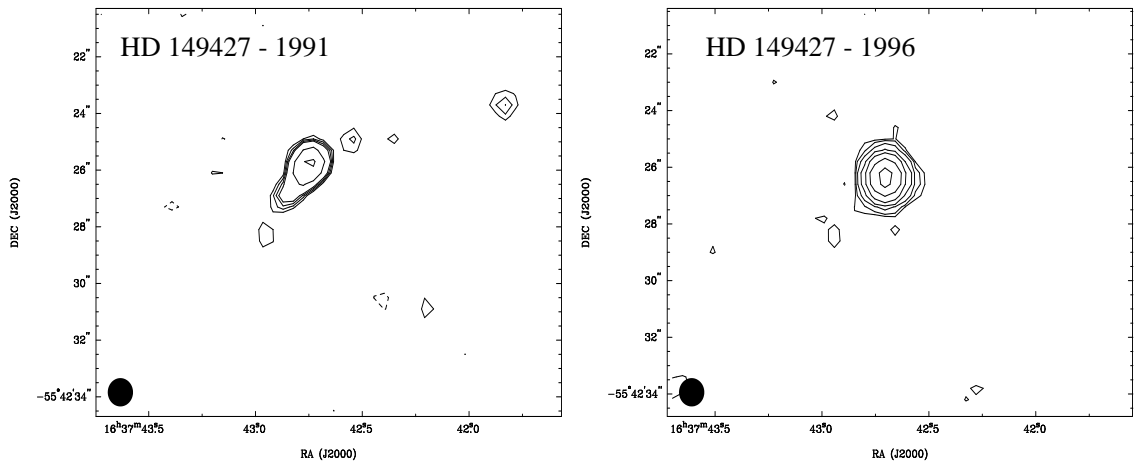,angle=0,width=15cm}
\caption{Radio maps of HD~149427 in 1991 and 1996 at 8.460 and 8.512 GHz respectively. Contours are plotted at $-$3, 3, 4, 5, 6, 12, 24 $\times\sigma$ for the 1991 map and at $-$3, 3, 6, 12, 24, 48, 96, 192 $\times\sigma$ for the 1996 map. See text for comparison with the optical images.}
\label{map} 
\end{center}
\end{figure*}

The radio counterpart to HD~149427 was detected during both epochs and at all frequencies with the exception of 1.344 GHz; Table~\ref{radio} lists the flux densities in each case, which were obtained by fitting a Gaussian to the centre of each image. Coverage of the $uv$ plane was insufficient for mapping at the lower frequencies but maps at 8.640 and 8.5056 GHz (1991 and 1996 respectively) are displayed in Fig.~\ref{map}. 

The 1991 image is clearly extended to the south-east at the $\sim 3-6\sigma$ level and an additional feature is seen to the north-west which is likely to be associated with the central source. A more distant feature is seen at approximately the position of point {\bf b}, but closer inspection suggests that there is a $1.7\arcsec$ (to milliarcsecond accuracy) difference in position between {\bf b} and this radio feature. The peak flux density of the 1996 image was twice that of 1991 but the map shows a single and much more point-like source. There is a possible extension to the west in the 1996 source but it is insufficiently resolved to determine whether it is real. The lack of discrete sources in the 1996 image would suggest that the extended emission in 1991 was indeed associated with HD~149427 and that the ejecta are relatively short-lived. A least-squares fit to the 1996 radio spectrum indicates a spectral index of 0.9, consistent with thermal emission from photoionised ejecta.

We suggest that the structure seen in 1991 may be the radio counterpart to the apparent sporadic ejecta seen to the north-west of the source in the optical images. We also note that the 1991 radio images were obtained just 3 months after the DSS2 image and so the presence of emission at the same position as point {\bf b} in the latter does not necessarily rule out a jet origin. Indeed, the presence of a variable radio source approximately coincident with the optical source {\bf b} increases the likelihood of association of the optical source {\bf b} with HD~149427. Again further observations, obtained simultaneously at optical and radio frequencies, would be extremely useful. 


\section{Conclusions}
We have presented the results of our {\sl HST}/WFPC2 imaging campaign for the symbiotic stars HD~149427 (PC 11), PU Vul, RT Ser, He2$-$104 (Southern Crab), V1329 Cyg (HBV 475), V417 Cen, AS 201, RS Oph and AG Peg. Extended emission has been detected around He2$-$104, V1329 Cyg and possibly HD~149427. In particular, comparison of our V1329 Cyg image with previously published {\sl HST}/FOC results indicates expanding ejecta, perhaps associated with an ejection event in 1982 ($\pm2$ years) at a velocity of $260\pm 50$ km~s$^{-1}$ in the plane of the sky and at an assumed distance of 3.4kpc. Radio images of HD~149427 confirm the presence of ejected material in 1991 along a similar axis to that of the possible extended optical emission in 1999; a 1996 radio observation was more point-like suggesting that any ejection events are relatively short-lived.

\section*{acknowledgements}
We are grateful to Augustin Skopal for useful discussions and to the referee, Romano Corradi for helpful comments on the original manuscript. The {\sl Hubble Space Telescope} is a joint project between the US National Aeronautics and Space Administration and the European Space Agency. The Australia Telescope is funded by the Commonwealth of Australia for operation as a National Facility managed by CSIRO. CB is supported by a PPARC Post Doctoral Research Assistantship.

\end{document}